\documentclass{aa9}
\usepackage[varg]{txfonts}
\usepackage{longtable}
\usepackage{lscape}
\usepackage{graphicx}

\newcommand{\kmprs}  {\mbox{\rm km\,s$^{-1}$}}
\newcommand{\feh} {\mbox{\rm [Fe/H]}}

\newcommand{\xfe} {\mbox{\rm [X/Fe]}}
\newcommand{\cfe} {\mbox{\rm [C/Fe]}}
\newcommand{\ofe} {\mbox{\rm [O/Fe]}}
\newcommand{\nafe} {\mbox{\rm [Na/Fe]}}
\newcommand{\mgfe} {\mbox{\rm [Mg/Fe]}}

\newcommand{\alfe} {\mbox{\rm [Al/Fe]}}
\newcommand{\almg} {\mbox{\rm [Al/Mg]}}
\newcommand{\sife} {\mbox{\rm [Si/Fe]}}
\newcommand{\sfe} {\mbox{\rm [S/Fe]}}
\newcommand{\cafe} {\mbox{\rm [Ca/Fe]}}
\newcommand{\scfe} {\mbox{\rm [Sc/Fe]}}
\newcommand{\scmg} {\mbox{\rm [Sc/Mg]}}
\newcommand{\tife} {\mbox{\rm [Ti/Fe]}}
\newcommand{\crfe} {\mbox{\rm [Cr/Fe]}}
\newcommand{\mnfe} {\mbox{\rm [Mn/Fe]}}
\newcommand{\nife} {\mbox{\rm [Ni/Fe]}}
\newcommand{\cufe} {\mbox{\rm [Cu/Fe]}}
\newcommand{\znfe} {\mbox{\rm [Zn/Fe]}}
\newcommand{\yfe} {\mbox{\rm [Y/Fe]}}
\newcommand{\zrfe} {\mbox{\rm [Zr/Fe]}}
\newcommand{\lafe} {\mbox{\rm [La/Fe]}}
\newcommand{\bafe} {\mbox{\rm [Ba/Fe]}}

\newcommand{\ymg} {\mbox{\rm [Y/Mg]}}
\newcommand{\yal} {\mbox{\rm [Y/Al]}}

\newcommand{\cuna} {\mbox{\rm [Cu/Na]}}

\newcommand{\alphafe} {\mbox{\rm [$\alpha$/Fe]}}
\newcommand{\teff}  {\mbox{$T_{\rm eff}$}}
\newcommand{\Tc}  {\mbox{$T_{\rm C}$}}

\newcommand{\logg}  {\mbox{{\rm log}\,$g$}}
\newcommand{\turb}  {\mbox{$\xi_{\rm turb}$}}

\newcommand{\ScII} {\ion{Sc}{ii}}

\newcommand{\CrI} {\ion{Cr}{i}}

\newcommand{\MnI} {\ion{Mn}{i}}
\newcommand{\FeI} {\ion{Fe}{i}}

\newcommand{\CuI} {\ion{Cu}{i}}

\newcommand{\BaII} {\ion{Ba}{ii}}

\def\ltsima{$\; \buildrel < \over \sim \;$}
\def\simlt{\lower.5ex\hbox{\ltsima}}
\def\gtsima{$\; \buildrel > \over \sim \;$}
\def\simgt{\lower.5ex\hbox{\gtsima}}
\begin{document}

\title{High-precision abundances of Sc, Mn, Cu, and Ba in solar twins}

\subtitle{Trends of element ratios with stellar age
\thanks{Based
on data products from observations made with ESO Telescopes
at the La Silla Paranal Observatory under programs
072.C-0488, 088.C-0323, 183.C-0972, 188.C-0265.}}

\author{P. E. Nissen}

\institute{Stellar Astrophysics Centre, 
Department of Physics and Astronomy, Aarhus University, Ny Munkegade 120, DK--8000
Aarhus C, Denmark.  \email{pen@phys.au.dk}.}

\date{Received 5 May 2016 / Accepted 15 June 2016}

\abstract
{}
{A previous study of correlations between element abundances and ages of
solar twin stars in the solar neighborhood is extended to include 
Sc, Mn, Cu, and Ba to obtain new 
information on the nucleosynthetic history of these elements.}
{HARPS spectra with $S/N\simgt600$ are used to derive
very precise ($\sigma\!\sim0.01$\,dex) differential abundances of Sc, Mn, Cu, and Ba
for 21 solar twins and the Sun. The analysis is based on
MARCS model atmospheres with parameters
determined from the excitation and ionization
balance of Fe lines. Stellar ages with internal errors less than 1\,Gyr are obtained by interpolation
in the \logg \,--\,\teff\ diagram between isochrones based on the Aarhus Stellar
Evolution Code.} 
{For stars younger than 6\,Gyr, \scfe , \mnfe , \cufe, and \bafe\ are tightly
correlated with stellar age, which is also the case for the other elements previously studied.
Linear relations between \xfe\ and age have $\chi^2_{\rm red} \! \sim \! 1$,
and for most stars the residuals do not depend on elemental condensation
temperature. 
For ages between 6 and 9\,Gyr, the \xfe \,-\,age correlations break down and the stars split
up into two groups having respectively high and low \xfe\ for the odd-$Z$ elements
Na, Al, Sc, and Cu.} 
{While stars in the solar neighborhood younger than $\sim \! 6$\,Gyr  
were formed from interstellar gas with a smooth chemical evolution,
older stars seem to have originated from regions enriched by supernovae with different
neutron excesses. Correlations between abundance ratios and stellar age suggest that:
i) Sc is made in Type II supernovae along with the $\alpha$-capture elements;
ii) the Type II to Ia yield ratio is
about the same for Mn and Fe; iii) Cu is mainly made by
the weak $s$-process in massive stars; iv) the Ba/Y yield ratio for  AGB stars 
increases with decreasing stellar mass; v) [Y/Mg] and 
[Y/Al] can be used as ``chemical clocks" when determining ages of solar metallicity stars.}


\maketitle

\section{Introduction}
\label{introduction} 
Precise determinations of stellar abundances by \citet{melendez09}
and \citet{ramirez09} have revealed significant variations of
abundance ratios among solar twin stars
that are correlated with elemental condensation 
temperature, \Tc\ \citep{lodders03}. Interestingly, the Sun
has a higher ratio between volatile and refractory elements
than the majority of solar twins, which could be caused by sequestration
of refractory elements in terrestrial planets in the solar system. 
If so, the slope of [X/Fe]\,\footnote{For two elements, X and Y,
with number densities $N_{\rm X}$ and $N_{\rm Y}$,
[X/Y] $\equiv {\rm log}(N_{\rm X}/N_{\rm Y})_{\rm star}\,\, - 
\,\,{\rm log}(N_{\rm X}/N_{\rm Y})_{\rm Sun}$.} versus \Tc\ may be used
as a signature of the occurrence of rocky planets around stars.
Alternatively, slope differences could be due to dust-gas separation
in connection with star formation \citep{onehag14, gaidos15}.

High-precision abundance studies based on HARPS spectra also
point to variations of the [X/Fe]\,-\,\Tc\ slope
among solar-type stars \citep{gonzalez.her10, gonzalez.her13,
adibekyan14}, but the slope does not seem to depend on
the occurrence of terrestrial planets. The same conclusion is reached by
\citet{schuler15} based on detailed abundances of seven Kepler stars
hosting small planets. Instead, \citet{adibekyan14},
\citet{maldonado15}, and \citet{maldonado16} find 
a correlation between the [X/Fe]\,-\,\Tc\ slope and stellar age. This is
confirmed by \citet[][hereafter Paper I]{nissen15}, who from a study of
precise abundances of 21 solar twins in the solar neighborhood
finds that ratios between volatile and
refractory elements, e.g., [C/Fe] and [O/Fe], are  correlated with 
stellar age.  The same result is obtained by \citet{spina16a} for 13 solar twins
belonging to the thin-disk population.
In both studies there are, however, indications of significant [X/Fe]\,-\,\Tc\ slope
differences at a given age suggesting that other processes than chemical
evolution play a role. This is supported by studies of stellar twins in wide 
binary systems, e.g., 16\,Cyg \citep{ramirez11, tuccimaia14}
and XO-2 \citep{teske15, biazzo15, ramirez15} for which there are
differences in \feh\ and the [X/Fe]\,-\,\Tc\ slope  between the components.

According to the results obtained in Paper\,I and by \citet{spina16a}, the
differences of abundance ratios among solar twin stars are mainly due to Galactic chemical
evolution.  The variation of the [X/Fe]\,-\,\Tc\ slope 
is a secondary effect that introduces some scatter in the [X/Fe]\,-\,age trends, but it 
does not prevent us from obtaining new knowledge about nucleosynthesis 
from the relations between abundance ratios and stellar age. Given that
the relative ages of solar twins can be determined with a precision
better than 1\,Gyr (see Paper I), we may in this way get more direct information on Galactic
chemical evolution than by studying the change of abundance ratios as a function
of metallicity, which method is hampered by a large scatter
in the age-metallicity relation \citep{edvardsson93, nordstrom04, haywood13}.

As shown in Paper I, the abundances of C, O, Mg, Al, and Zn 
relative to Fe for solar twins belonging to the thin-disk
population increase by
0.10 to 0.15\,dex as the stellar age increases from $\sim \! 1$ to 8\,Gyr.
This can be explained if C, O, Mg, Al, and Zn are  mainly made in 
high-mass stars exploding as 
Type II supernovae (SNe) on a relatively short time scale,
whereas Type Ia SNe provide an important contribution to Fe  on a longer time scale.
\sife , \sfe , and \tife\ change in the same way as a function of age
but with an amplitude of about 0.05\,dex only. \cafe\ on the other hand
is nearly the same for all ages with a scatter of 0.01\,dex only, which
is a surprise because Ca is normally considered to be an $\alpha$-capture
element mainly produced in Type II SNe like Mg, Si, and S. \nafe\ and \nife\
show considerable scatter among the older stars but are tightly
correlated with each other. Furthermore, \yfe\ increases steeply
with decreasing stellar age, which can be explained by an increasing
contribution to the abundance of $s$-process elements from low-mass
AGB stars in the course of time \citep{travaglio04}.

In the present paper, the study of solar twins in Paper I is extended to
include Sc, Mn, Cu, and Ba. The origin and nucleosynthesis of these elements
have been much discussed \citep[e.g.,][]{romano10, serminato09}, which to some extent is
due to uncertainties in the metallicity and age trends.
In some works \citep{gratton91, prochaska00a}, the Sc/Fe ratio 
is found to be close to solar at all metallicities, whereas other studies
\citep{zhao90, nissen00, reddy06, adibekyan12, battistini15} point to
enhanced Sc abundances (relative to Fe) in metal-poor thick-disk stars
similar to the enhancement of $\alpha$-elements.
Mn and Cu are found to be under-abundant relative to Fe for metal-poor disk and halo stars
\citep{sneden91,mishenina02,reddy06, feltzing07, nissen11, adibekyan12, battistini15, mishenina15a} 
if spectral lines are analyzed assuming local thermodynamic equilibrium (LTE).
Corrections for non-LTE effects \citep{bergemann08, yan15, yan16} modify,
however, the metallicity trends of \mnfe\ and \cufe\ very significantly.
In the case of Ba, some investigations point to enhanced \bafe\
values in young stars \citep{edvardsson93, bensby07}. This is, however,
not confirmed by \citet{mishenina13}.

In the following Sect. 2, the stellar parameters and abundances derived in Paper\,I
are reviewed,  and ages based on the Aarhus Stellar
Evolution Code are determined and compared to the ages derived in Paper\,I
from Yonsei-Yale (YY) isochrones. The determination of Sc, Mn, Cu, and Ba abundances
is presented in Sect. 3. Trends of abundance ratios as a function of stellar age
and elemental condensation temperature are discussed in Sect. 4 together with
some considerations about the nucleosynthesis of odd-$Z$ and neutron-capture 
elements and the use of \ymg\ or \yal\ as ``chemical clocks" for determining
stellar ages.  A summary and some conclusions are given in Sect. 5.

\section{Stellar parameters and ages}
\label{parameters}

The solar twins were selected from 
the analysis of HARPS spectra \citep{mayor03} by \citet{sousa08}
to have atmospheric parameters that agree with those of the Sun within 
$\pm 100$\,K in effective temperature, \teff , $\pm 0.15$\,dex
in logarithmic surface gravity, \logg , and $\pm 0.10$\,dex in metallicity, \feh .
The range in \logg\ is somewhat larger than the range ($\pm 0.10$\,dex) 
adopted by \citet{ramirez09} in their definition of a solar twin. It is also noted
that none of the stars is a ``perfect good solar twin" \citep{cayrel96}
in the sense that it has the same mass, age, and composition as the 
Sun within the estimated errors. The stars 
range in age from 0.7\,Gyr to about 9\,Gyr and show a diversity of
chemical compositions including large differences in lithium abundance
correlated with stellar age \citep{carlos16}.

As an additional selection criteria, only stars having 
HARPS spectra with a signal-to-noise ratio $S/N \ge 600$
were included. This resulted in a list of 21 solar twins, three of
which have enhanced \alphafe\ ratios and probably belong to the thick-disk
population \citep{haywood13} or a special population of high-$\alpha$ metal-rich
(h$\alpha$mr) stars \citep{adibekyan11}. The other 18 stars have typical thin-disk 
abundances and kinematics. All stars have distances less than 60\,pc
according to their Hipparcos parallaxes \citep{vanleeuwen07}.

The extremely high S/N and resolution ($R \simeq 115\,000$)
of the HARPS spectra make it possible to determine very precise
parameters and abundances of the solar twins relative to the Sun.
The analysis in Paper\,I was made differentially, line-by-line, relative to
a HARPS solar flux spectrum with $S/N \sim 1200$  observed in reflected 
sunlight from the asteroid Vesta. \teff\ and \logg\
were determined by requesting that the iron abundances derived have no systematic
dependence on excitation and ionization potential of the Fe lines applied.
The estimated errors of the parameters are $\sigma (\teff ) = \pm 6$\,K and
$\sigma (\logg ) = \pm 0.012$\,dex, and abundances of C, O, Na, Mg, Al,
Si, S, Ca, Ti, Cr, Fe, Ni, Zn, and Y were determined with a
typical precision of $\pm 0.01$\,dex. 

By interpolating to the observed values of \teff , \logg , \feh ,
and \alphafe\ in a set of YY isochrones \citep{yi01, kim02}, stellar ages 
were derived in Paper\,I\footnote{The YY isochrones were offset by +0.005\,dex
in \logg\ to get an age of 4.55\,Gyr for the Sun.}. 
An internal age precision of 0.4 to 0.8\,Gyr was estimated,
but due to  possible systematic errors in the model calculations,  
the ages of the youngest and oldest stars
may be more uncertain. In order to test this,
the Aarhus Stellar Evolution Code (ASTEC) described by \citet{jcd08} has
been used to calculate isochrones in the \teff \,-\,\logg\ plane for a range 
of compositions covering those of the solar twins.  In contrast to the
YY models, the ASTEC models include  diffusion and settling 
of heavy elements \citep{jcd93}. Therefore, the ASTEC ages were obtained by
interpolating in the surface heavy element abundance ($Z_s$) for the
isochrones to the observed Z-value of the star. 

\begin{figure}
\resizebox{\hsize}{!}{\includegraphics{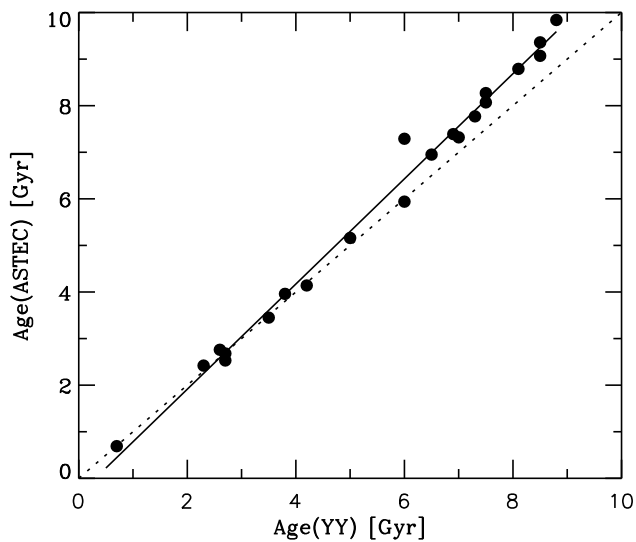}}
\caption{Comparison of stellar ages derived from YY and ASTEC isochrones.
The line shows the linear fit to the data given by Eq. (1). The dotted
line shows the 1:1 relation.}
\label{fig:ages}
\end{figure}

As seen from Fig. \ref{fig:ages}, the ASTEC ages  of the oldest stars
are systematically higher than the ages derived from YY isochrones.
A linear fit to the data gives 
\begin{eqnarray}
Age \, {\rm (ASTEC)} = -0.3 \, (\pm 0.1) + 1.13 \, (\pm 0.02) \,\, Age\,{\rm (YY) \, [Gyr]} 
\end{eqnarray}
with a standard deviation of 0.28\,Gyr corresponding to 0.20\,Gyr on each of the 
two sets of ages.
The residuals in the fit are to some degree correlated with stellar metallicity,
which may be due to different assumptions about the change of helium
abundance as a function of heavy element abundance. For the ASTEC models
$\Delta Y / \Delta Z = 1.4$ is adopted, whereas the YY models are based on
$\Delta Y / \Delta Z = 2.0$.  In any case, the scatter in the relation between
the two sets of ages
is smaller than the age errors (0.4 to 0.8\,Gyr) arising from the 
uncertainties in \teff\ and \logg . Hence, the use of a different sets 
of isochrones has only a small effect when comparing stars of similar age, but
it affects the age scale; for the oldest stars, 10\,\% higher ages are derived
when applying ASTEC instead of YY isochrones.

The YY and ASTEC models are based on similar
prescriptions for energy generation, opacity and equations of state,
and are calibrated on the Sun adopting a mass ratio between heavy 
elements and hydrogen of ($Z/X)_{\odot}$\,=\,0.0244 at the surface 
\citep{grevesse96}. Both sets of models include 
helium diffusion, but as mentioned above only the ASTEC models include diffusion of heavy 
elements, which means that the opacity in the interior of an ASTEC model
of an old star is higher than in a YY model with the same surface abundance.
This may be the reason for the different age scales.

As a further test of systematic errors in the ages, ASTEC isochrones
have been compared to the often used
Padova\,\footnote{{\tt http://stev.oapd.inaf.it/cgi-bin/cmd}}
(PARSEC) isochrones \citep{bressan12}, 
which are also based on stellar models including diffusion of 
heavy elements. There is an excellent agreement between the two sets
of isochrones at solar heavy element abundance for the \teff \,-\,\logg\ ranges of 
solar twins suggesting that the ASTEC and PARSEC age scales are about the same.  

Considering that the inclusion of heavy element diffusion makes the
ASTEC models more realistic than the YY models,
stellar ages derived from the ASTEC isochrones will be used in the present paper.
The relations between abundance ratios and stellar age given in Paper\,I
may be converted to the ASTEC age scale by using Eq. (1).

\section{Sc, Mn, Cu, and Ba abundances}
\label{abundances}
\begin{figure*}
\resizebox{\hsize}{!}{\includegraphics{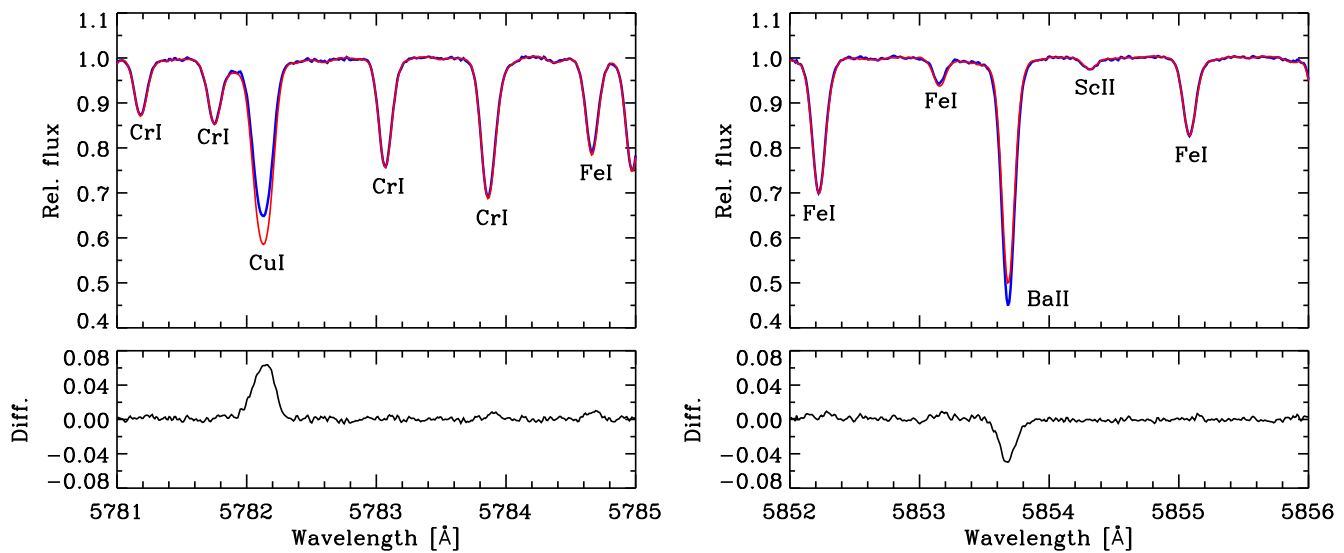}}
\caption{The solar flux HARPS spectrum 
(red line) in comparison with the HARPS spectrum of \object{HD\,96116} (blue line)
near the \CuI\ line at 5782.12\,\AA\ (left panel) and the \BaII\ line at 5853.70\,\AA\
(right panel).
The lower panels show the difference (HD\,96116 -- Sun) between the two spectra.}
\label{fig:5782-5853}
\end{figure*}

Sc, Mn, Cu, and Ba abundances are determined from the spectral
lines listed in Table \ref{table:linedata}. The lines selected 
have nearby ``continuum" regions almost free of lines. 
Equivalent widths (EWs) were measured by Gaussian fitting using 
the IRAF {\tt splot} task except for
the \MnI\ 5394.69\,\AA\ line that has a ``boxlike" profile
due to hyperfine splitting and was measured
by direct integration\,\footnote{The magnetic sensitive \MnI\ 5394.69\,\AA\ line
in the solar flux spectrum has been found to vary in strength over the solar
cycle \citep{danilovic16}, but the Mn abundances of the solar twins
derived from this line show no excessive variation relative to the Mn abundances
derived from the other \MnI\ lines}. Care was taken to use the same continuum windows
in all stars. To illustrate the high quality of the spectra, Fig. 
\ref{fig:5782-5853} shows a comparison of 
the spectrum of HD\,96116 with the solar spectrum for the regions of the 
\CuI\  5782.12\,\AA\ and \BaII\ 5853.70\,\AA\ lines. HD\,96116 is the
youngest star in the sample (age $\simeq 0.7$\,Gyr)
and has a somewhat higher temperature ($\teff = 5846$\,K)
and surface gravity ($\logg = 4.50$) but almost he same metallicity
($\feh = 0.006$) as the Sun. As seen, the \CrI\ and \FeI\ lines have nearly the
same strengths in the two spectra, whereas the \CuI\ line is weaker and the 
\BaII\ line is stronger in the spectrum of HD\,96116. As we shall see
in Sect. 4, this is related to the different trends of \cufe\ and
\bafe\ as a function of stellar age.

\begin{table}
\caption[ ]{Line data and equivalent widths measured for the HARPS
solar flux spectrum.}
\label{table:linedata}
\setlength{\tabcolsep}{0.30cm}
\begin{tabular}{cccc}
\noalign{\smallskip}
\hline\hline
\noalign{\smallskip}
  ID & Wavelength & $\chi_{\rm exc}$  & $EW_{\odot}$ \\
          &  [\AA ]    &  [eV]  &    [m\AA ]   \\
\noalign{\smallskip}
\hline
\noalign{\smallskip}
\ScII\ & 5684.20 & 1.507 &   37.1  \\
\ScII\ & 6245.65 & 1.507 &   33.1  \\
\MnI\  & 5004.89 & 2.920 &   14.0  \\
\MnI\  & 5394.69 & 0.000 &   79.7  \\
\MnI\  & 5399.50 & 3.850 &   39.1  \\
\MnI\  & 5432.55 & 0.000 &   51.7  \\
\MnI\  & 6013.52 & 3.073 &   87.3  \\
\MnI\  & 6016.67 & 3.075 &   96.6  \\
\MnI\  & 6021.81 & 3.075 &   95.9  \\
\CuI\  & 5218.20 & 3.820 &   53.6  \\
\CuI\  & 5782.12 & 1.642 &   79.2  \\
\BaII\ & 5853.70 & 0.604 &   63.9  \\
\BaII\ & 6496.90 & 0.604 &  100.3  \\
\noalign{\smallskip}
\hline
\end{tabular}

\end{table}

The set of MARCS model atmospheres \citep{gustafsson08} described
in Paper\,I  and the Uppsala BSYN program were used to
calculate equivalent widths as a function of element abundance
assuming LTE. Interpolation to the observed EW then provides the stellar
abundance. Hyperfine splitting with data from
\citet{prochaska00b} was taken into account as described by \citet{nissen11}.
Microturbulence velocities were adopted from Paper\,I, and
collisional broadening was included in the \citet{unsold55} approximation
using an enhancement factor of two.

Abundance errors arising from the EW measurements are estimated by comparing
abundances derived from the individual lines. Given that a homogeneous set of
spectra has been used, we may calculate an average standard deviation of \xfe\
\begin{eqnarray}
<\sigma \xfe > \, = \, \sqrt{\sum_{i=1}^{N_{\rm stars}} \sigma_i^2 \xfe \,/\, N_{\rm stars} }
\end{eqnarray}
where
\begin{eqnarray}
\sigma_i^2 \xfe \, = \, \sum_{k=1}^{N_{\rm lines}} (\xfe _{k,i} - <\xfe _{i}>)^2 \, /\, (N_{\rm lines}-1)
\end{eqnarray}
is the variance for a given star. The corresponding average standard error
of the mean value of \xfe\ 
\begin{eqnarray}
\sigma _{\rm EW} \, = \, <\sigma \xfe > / \sqrt{N_{\rm lines}} 
\end{eqnarray}
is given in column 3 of Table \ref{table:errors}.
This error is only 0.003\,dex in the case of Mn that has 
seven lines available,  but is higher for elements with
abundances based on two lines.

In addition to the error arising from the EW measurements, the uncertainties in the
atmospheric parameters contribute to the total error of the abundances.
For some abundance ratios, this error is small, because the derived 
abundances of two elements may have approximately the same dependence on 
\teff , \logg , \feh , and microturbulence.  
As a continuation of Table 4 in Paper\,I, the two sets of errors are
given in Table \ref{table:errors} for the abundance ratios
discussed in the present paper.

\begin{table}
\caption[ ]{Errors of abundance ratios.}
\label{table:errors}
\centering
\setlength{\tabcolsep}{0.40cm}
\begin{tabular}{crrr}
\noalign{\smallskip}
\hline\hline
\noalign{\smallskip}
       & $\sigma_{\rm atm.par.}$ & $\sigma_{\rm EW}$ & $\sigma_{\rm adopted}$ \\
       &  [dex] &  [dex] &  [dex]  \\
\noalign{\smallskip}
\hline
\noalign{\smallskip}
 \scfe  &   0.007 & 0.006 &  0.010 \\
 \mnfe  &   0.002 & 0.003 &  0.006 \\
 \cufe  &   0.004 & 0.006 &  0.009 \\
 \bafe  &   0.008 & 0.011 &  0.014 \\
 \scmg  &   0.008 & 0.011 &  0.014 \\
 \cuna  &   0.005 & 0.007 &  0.010 \\
 \ymg   &   0.008 & 0.011 &  0.014 \\
 \yal   &   0.008 & 0.008 &  0.012 \\
 \almg  &   0.003 & 0.012 &  0.013 \\
\noalign{\smallskip}
\hline
\end{tabular}
\end{table}

As the solar twins have similar atmospheric parameters, differential non-LTE
corrections are expected to be small. Still, there could be spurious trends
of abundance ratios as a function of \teff , \logg , and \feh , if non-LTE effects
are not taken into account. Thus, \citet{bergemann08} found that the non-LTE
correction for Mn increases with decreasing metallicity.
Using the same model atom,  M. Bergemann (private communication) has
calculated non-LTE  corrections for the solar twins in this paper.
The amplitude of the corrections is $\pm 0.01$\,dex over the metallicity range 
from $\feh = -0.10$ to +0.10\,dex and have
been applied although the effect on the \mnfe -age relation is marginal.

In the case of Cu, \citet{yan15, yan16} have derived non-LTE corrections that increase
with decreasing metallicity in the same way as for Mn. 
Based on the corrections for the 5218.2 and 5782.1\,\AA\ lines 
\citep[][Table 2]{yan15},
it is possible to estimate the change of non-LTE corrections
relative to those of the Sun as a function of \teff , \logg , and \feh .
Applying these relations to the solar twins results in 
differential non-LTE corrections reaching 
+0.015\,dex for stars having $\feh \sim -0.1$ and low gravity. Hence, the
corrections are of some importance for the age trends of Cu and have been
taken into account. The same method has been applied for Ba adopting non-LTE
corrections calculated by \citet{korotin15} for the 5853.7 and 6496.9\,\AA\ \BaII\
lines. This results in rather small differential corrections for the solar twins
reaching an extreme of $-0.007$\,dex at the lowest gravities. In the case of Sc,
non-LTE effects are very small for \ScII\ lines \citep{zhang08, zhang14},  
so differential corrections for solar twins have not been applied.

In Paper\,I, non-LTE corrections \citep{lind11, lind12}
were applied for Na and Fe. For the other elements
studied, differential non-LTE effects 
were either estimated to be negligible or non-LTE calculations were not available.
Furthermore, all non-LTE calculations are
based on 1D model atmospheres; 3D non-LTE calculations may give somewhat different
results. To take this additional uncertainty into account, an error of 0.005\,dex
as in Paper\,I
has been added in quadrature to the errors arising from the uncertainties in the
atmospheric parameters and the EW measurements. This leads to the adopted errors of
\xfe\ given in the last column of Table \ref{table:errors}.

\section{Results and discussion}
\label{discussion}
The derived values of \scfe , \mnfe , \cufe , and \bafe\ are given in Table \ref{table:abun}
together with the atmospheric parameters and ages derived from the ASTEC isochrones.
Masses determined from the ASTEC evolutionary tracks are also included. They compare
very well with the masses determined from Yonsei-Yale stellar evolutionary tracks
\citep{yi03} given in Paper\,I; the average deviation (YY -- ASTEC) is 0.005\,$M_{\odot}$ 
with a rms deviation of 0.010\,$M_{\odot}$ suggesting that the masses have a precision
of about 0.01\,$M_{\odot}$. 

In the following subsections, the new data on Sc, Mn, Cu, and Ba are discussed together
with the abundances for C, O, Na, Mg, Al, Si, S, Ca, Ti, Cr, Fe, Ni, Zn,  and Y 
derived in Paper\,I.

\begin{table*}
\caption[ ]{Stellar atmospheric parameters, ages, masses, and abundance ratios.}
\label{table:abun}
\setlength{\tabcolsep}{0.20cm}
\begin{tabular}{lccrccccrrrr}
\noalign{\smallskip}
\hline\hline
\noalign{\smallskip}
  Star & \teff & \logg & \feh & \turb & Age\tablefootmark{a} &  $\sigma$(Age) & Mass\tablefootmark{a} & \scfe & \mnfe & \cufe & \bafe  \\
  & [K] &  &  & [\kmprs ] & [Gyr] & [Gyr] & [$M_{\odot}$]  & &  &  &  \\
\noalign{\smallskip}
\hline
\noalign{\smallskip}
HD\,2071 &   5724 & 4.490 & $-$0.084 &   0.96 &    3.5 &  0.8 & 0.97 &  0.000 & $-$0.014 & $-$0.036 &  0.081 \\
HD\,8406 &   5730 & 4.479 & $-$0.105 &   0.95 &    4.1 &  0.8 & 0.96 & $-$0.009 & $-$0.025 & $-$0.064 &  0.123 \\
HD\,20782 &   5776 & 4.345 & $-$0.058 &   1.04 &    8.1 &  0.4 & 0.97 &  0.005 & $-$0.047 & $-$0.036 & $-$0.031 \\
HD\,27063 &   5779 & 4.469 &  0.064 &   0.99 &    2.8 &  0.6 & 1.03 & $-$0.026 & $-$0.015 & $-$0.052 &  0.090 \\
HD\,28471 &   5754 & 4.380 & $-$0.054 &   1.02 &    7.3 &  0.4 & 0.97 &  0.063 & $-$0.032 &  0.035 & $-$0.010 \\
HD\,38277 &   5860 & 4.270 & $-$0.070 &   1.17 &    7.8 &  0.4 & 1.01 &  0.026 & $-$0.053 & $-$0.015 & $-$0.032 \\
HD\,45184 &   5871 & 4.445 &  0.047 &   1.06 &    2.7 &  0.5 & 1.05 &  0.002 & $-$0.023 & $-$0.025 &  0.060 \\
HD\,45289\tablefootmark{b} &   5718 & 4.284 & $-$0.020 &   1.06 &    9.4 &  0.4 & 0.99 &  0.110 & $-$0.055 &  0.060 & $-$0.039 \\
HD\,71334 &   5701 & 4.374 & $-$0.075 &   0.98 &    8.8 &  0.4 & 0.94 &  0.049 & $-$0.028 &  0.003 & $-$0.027 \\
HD\,78429 &   5756 & 4.272 &  0.078 &   1.05 &    8.3 &  0.4 & 1.02 &  0.007 & $-$0.043 &  0.014 & $-$0.006 \\
HD\,88084 &   5768 & 4.424 & $-$0.091 &   1.02 &    5.9 &  0.6 & 0.97 &  0.063 & $-$0.014 &  0.027 & $-$0.013 \\
HD\,92719 &   5813 & 4.488 & $-$0.112 &   1.00 &    2.5 &  0.6 &  0.99 & 0.009 & $-$0.036 & $-$0.048 &  0.142 \\
HD\,96116 &   5846 & 4.503 &  0.006 &   1.02 &    0.7 &  0.7 & 1.04 & $-$0.055 & $-$0.027 & $-$0.098 &  0.190 \\
HD\,96423 &   5714 & 4.359 &  0.113 &   0.99 &    7.3 &  0.6 & 1.01 &  0.038 & $-$0.020 &  0.037 & $-$0.036 \\
HD\,134664 &   5853 & 4.452 &  0.093 &   1.01 &    2.4 &  0.5 & 1.06 & $-$0.023 & $-$0.028 & $-$0.040 &  0.085 \\
HD\,146233 &   5809 & 4.434 &  0.046 &   1.02 &    4.0 &  0.5 & 1.03 & $-$0.001 & $-$0.016 & $-$0.021 &  0.064 \\
HD\,183658 &   5809 & 4.402 &  0.035 &   1.04 &    5.2 &  0.5 & 1.02 & 0.026 &  0.006 &  0.036 & $-$0.028 \\
HD\,208704 &   5828 & 4.346 & $-$0.091 &   1.08 &    7.4 &  0.4 &  0.98 & 0.029 & $-$0.043 & $-$0.012 & $-$0.011 \\
HD\,210918\tablefootmark{b} &   5748 & 4.319 & $-$0.095 &   1.07 &    9.1 &  0.4 & 0.97 &  0.076 & $-$0.070 &  0.007 & $-$0.006 \\
HD\,220507\tablefootmark{b} &   5690 & 4.247 &  0.013 &   1.07 &    9.8 &  0.4 & 1.00 &  0.115 & $-$0.053 &  0.094 & $-$0.045 \\
HD\,222582 &   5784 & 4.361 & $-$0.004 &   1.07 &    7.0 &  0.4 & 1.00 &  0.058 & $-$0.019 &  0.059 & $-$0.024 \\
\noalign{\smallskip}
\hline
\end{tabular}
\tablefoot{
\tablefoottext{a}{The age and mass are derived from ASTEC models}.
\tablefoottext{b}{$\alpha$-enhanced star}.
}
\end{table*}

\subsection{Age correlations}
\label{age-correlations}
\begin{figure*} \resizebox{\hsize}{!}{\includegraphics{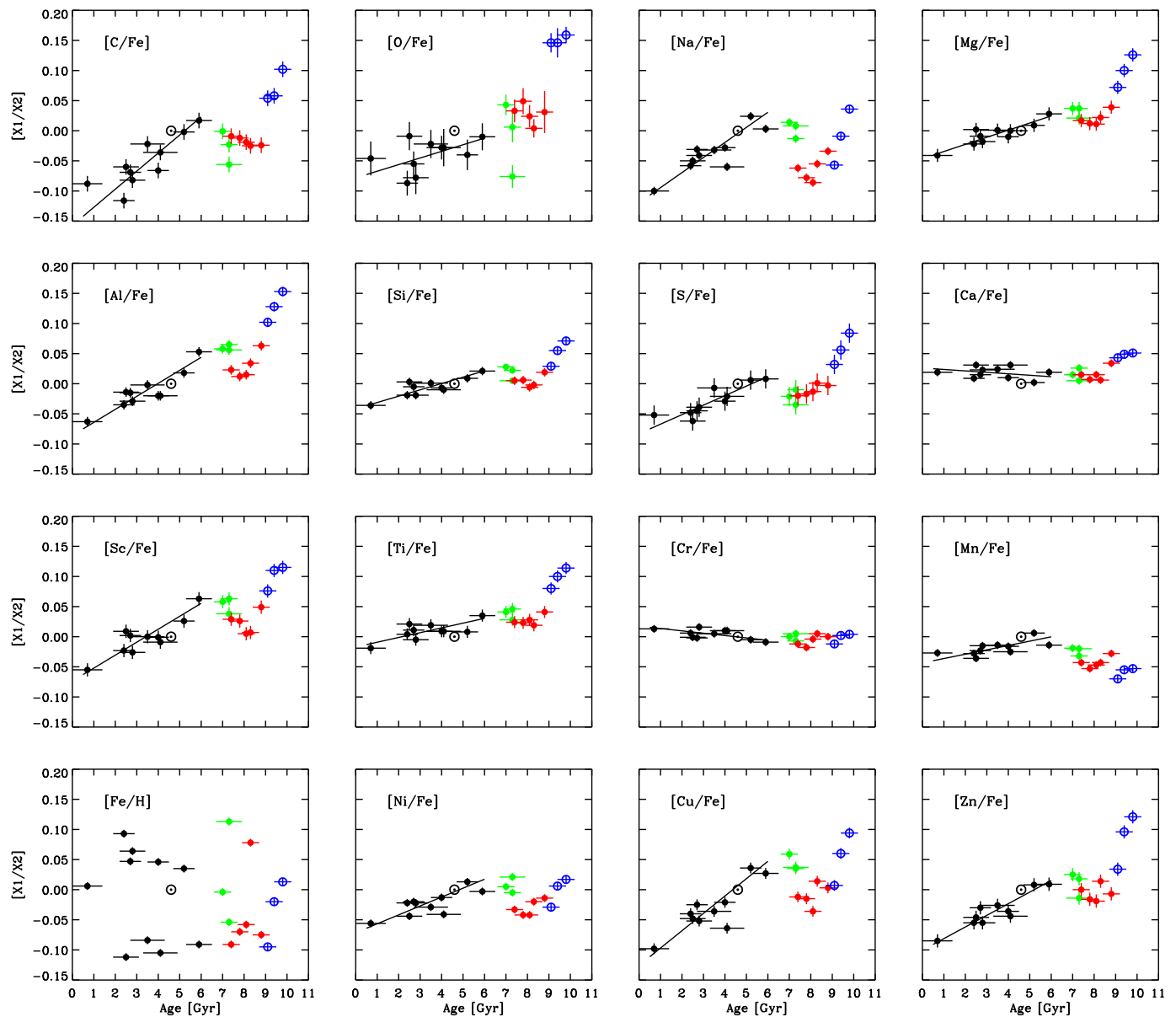}}
\caption{Abundance ratios as a function of stellar (ASTEC) age. For stars younger than 6 Gyr
(black filled circles) the lines show linear fits to the data. The Sun, shown with the
$\odot$ symbol, was not included in the fits. Red filled
circles show five old stars with low \nafe\ relative to three stars with similar age
shown with green filled circles. Three \alphafe -enhanced stars are shown with open blue circles.}
\label{fig:abun-age}
\end{figure*}

\begin{figure}
\resizebox{\hsize}{!}{\includegraphics{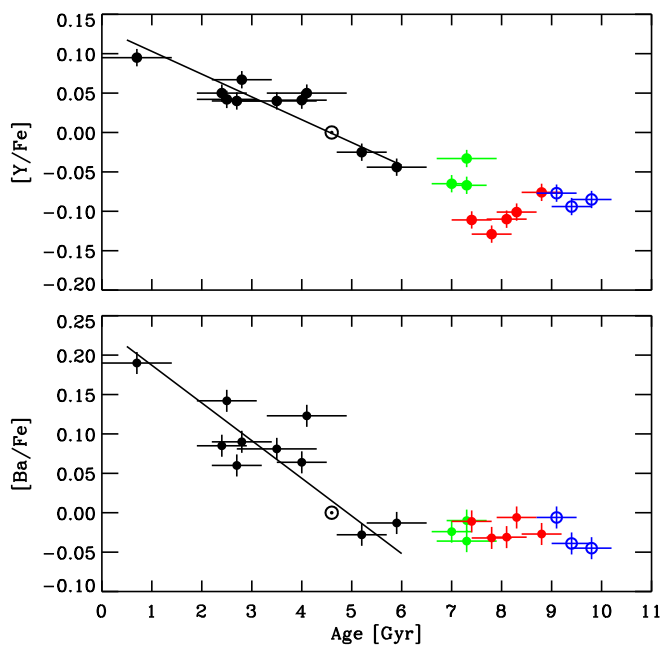}}
\caption{\yfe\ and \bafe\ as a function of stellar age with the
same symbols as in Fig. \ref{fig:abun-age}.}
\label{fig:bay-age}
\end{figure}

Figures \ref{fig:abun-age} and \ref{fig:bay-age} show abundances  
relative to Fe as a function of stellar age. As seen, all ratios 
are tightly correlated with age for stars younger than 6\,Gyr,
but the \xfe -age relations of odd-$Z$ elements and Ni tend to break down for
the older stars. 
This is mainly due to a group of five stars marked in red, which have
low values of \nafe , \alfe , \mnfe , \scfe , \nife , and \cufe\
relative to the group of three stars marked in green
(to be named as low- and high-\nafe\ stars, respectively).

Linear fits ($\xfe = a + b \,\cdot\,Age$) for the 10 stars younger than  6\,Gyr
were determined by a maximum likelihood program that includes errors in both
coordinates \citep{press92}. The coefficients $a$ and $b$
are given in Table \ref{table:fits} together with the reduced chi-squares,
which are satisfactory close to one.
For most elements, the \xfe -age slopes 
are steeper than those given in Paper\,I, where all stars except the three
$\alpha$-enhanced were included in the age fits.
These differences in slope arise because \xfe\ for stars older than 6\,Gyr tend
to lie below the fits obtained for stars younger than 6\,Gyr,
not only for the odd-$Z$ elements and Ni but also in the cases of C, Si, S, and Zn.

\begin{table}
\caption[ ]{Linear fits of \xfe\ as a function of age for stars
younger than 6\,Gyr, i.e., without including the Sun.}
\label{table:fits}
\centering
\setlength{\tabcolsep}{0.10cm}
\begin{tabular}{rcccc}
\noalign{\smallskip}
\hline\hline
\noalign{\smallskip}
       & $a$   & $b$                        & $\sigma \xfe$\,\tablefootmark{a}  & $\chi^2_{\rm red}$ \\
       & [dex] & $10^{-3}$\,dex\,Gyr$^{-1}$ &         [dex] &                    \\
\noalign{\smallskip}
\hline
\noalign{\smallskip}
 \cfe  &  $-0.157\,\pm 0.017$ & $+30.0\,\pm 4.6$ & 0.026 & 1.4 \\
 \ofe  &  $-0.078\,\pm 0.017$ & $+11.0\,\pm 4.5$ & 0.026 & 1.1 \\
 \nafe &  $-0.119\,\pm 0.011$ & $+25.0\,\pm 3.1$ & 0.020 & 1.1 \\
 \mgfe &  $-0.046\,\pm 0.005$ & $+11.8\,\pm 1.4$ & 0.009 & 0.5  \\
 \alfe &  $-0.086\,\pm 0.009$ & $+21.6\,\pm 2.1$ & 0.015 & 0.9  \\
 \sife &  $-0.042\,\pm 0.006$ & $+10.5\,\pm 1.6$ & 0.009 & 1.0  \\
 \sfe  &  $-0.083\,\pm 0.008$ & $+15.8\,\pm 2.1$ & 0.013 & 0.5  \\
 \cafe &  $+0.026\,\pm 0.007$ & $ -2.4\,\pm 2.1$ & 0.010 & 2.6  \\
 \scfe  &  $-0.074\,\pm 0.010$ & $+21.5\,\pm 2.9$ & 0.017 & 1.1  \\
 \tife &  $-0.017\,\pm 0.007$ & $ +7.9\,\pm 1.8$ & 0.011 & 1.0  \\
 \crfe &  $+0.018\,\pm 0.004$ & $ -3.9\,\pm 1.2$ & 0.007 & 1.2  \\
 \mnfe &  $-0.043\,\pm 0.007$ & $ +7.2\,\pm 2.0$ & 0.010 & 1.3  \\
 \nife &  $-0.072\,\pm 0.010$ & $+14.8\,\pm 2.7$ & 0.015 & 1.8  \\
 \cufe &  $-0.126\,\pm 0.013$ & $+28.8\,\pm 3.6$ & 0.024 & 1.1  \\
 \znfe &  $-0.101\,\pm 0.007$ & $+19.3\,\pm 2.1$ & 0.013 & 0.6  \\
 \yfe  &  $+0.132\,\pm 0.012$ & $-28.9\,\pm 3.3$ & 0.020 & 0.9  \\
 \bafe &  $+0.235\,\pm 0.024$ & $-47.8\,\pm 6.7$ & 0.041 & 1.5  \\
\noalign{\smallskip}
\hline
\end{tabular}
\tablefoot{
\tablefoottext{a}{Standard deviation of \xfe\ for the linear fit}.}
\end{table}

Interestingly, there are also differences in the kinematics of
the groups of stars shown in Fig. \ref{fig:abun-age}.
According to data from the Geneva-Copenhagen survey
\citep{nordstrom04, holmberg09}, the ten stars younger than 6\,Gyr
have about the same mean Galactocentric distance\footnote{$R_{\rm m}$
is defined as the mean value of the peri- and apo-galactocentric distances},
$R_{\rm m}$, in their orbits. The average value is $< \! R_{\rm m} \! >$\, = 8.1\,kpc
(i.e., close to the mean Galactocentric distance of the Sun, $R_{\rm m, \odot} \simeq 8.2$\,kpc)
with a rms dispersion of $\pm 0.4$\,kpc.
Assuming that $R_{\rm m}$ is a proxy of
the Galactocentric distance of the stellar birthplace
\citep{wielen77, grenon87, edvardsson93} despite of the possible effect
of radial migration \citep{sellwood02, schonrich09, gustafsson16},
it follows that the solar twins younger than 6\,Gyr
and the Sun were born within a narrow range in Galactocentric distance.
This makes it understandable that these stars originate from interstellar gas
with a good mixing of SNe products causing abundance ratios to develop in a
smooth and homogeneous way. For solar twins older than 6\,Gyr,
the dispersion in $R_{\rm m}$ is larger. The three high-Na stars
have $< \! R_{\rm m} \! > \, = \, 7.5 \pm 0.7$\,kpc and the five low-Na stars
have  $< \! R_{\rm m} \! > \, = \, 7.2 \pm 0.8$\,kpc. Hence, the differences
of abundance ratios among stars with ages between 6 and 9\,Gyr
may be related to spatial inhomogeneities in the chemical evolution of 
the early Galactic disk.

The differences in \xfe\ for Na, Al, Sc, Mn, Ni, and Cu for stars in the
6 to 9\,Gyr age range  cannot be explained
by spacial variations in the ratio between Type II and Type Ia SNe.
Such variations would also affect \xfe\ for the $\alpha$-elements, but
as seen from Fig. \ref{fig:abun-age}, there is no or only a very small
difference in \ofe , \mgfe , \sife , \sfe , \cafe , and \tife\ between the
high- and low-\nafe\ stars. Instead, the variations in \xfe\ for odd-$Z$ 
elements may be connected to a dependence of Type II SNe yields
on heavy element abundance (especially C and O)  
of the progenitors \citep[][Table 3]{kobayashi06}. 
This dependence arises because the production of the odd-$Z$
elements in core-collapse SNe is affected by the neutron excess, which is controlled
by the $^{22}$Ne\,$(\alpha , n)\,^{25}$Mg reaction,
where $^{22}$Ne comes from double $\alpha$-capture on $^{14}$N made in CNO burning at
the expense of C and O. Hence, one could imagine that the low-\nafe\ stars were formed in
regions having a lower heavy-element abundance than the regions where the 
high-\nafe\ stars were formed. Furthermore, the correlation
between variations of \nife\ and \nafe\ may be explained if the
yield of the dominating isotope, $^{58}$Ni, depends on the neutron excess as suggested
by \citet{venn04}.

There is no significant trend of \xfe\ as a function of \feh \ for Sc, Mn, Cu, and Ba.
Excluding the three $\alpha$-enhanced stars, the average rms scatter at a given \feh\ is 
$\sigma \scfe \, = \, 0.033$, $\sigma \mnfe \, = \, 0.014$, 
$\sigma \cufe \, = \, 0.042$, and $\sigma \bafe \, = \, 0.071$\,dex.
This lack of correlation between \xfe\ and  \feh\ is related to the fact
that there is no correlation between metallicity and age for the solar twins, but
a very significant dispersion ($\sigma \feh \sim \pm 0.07$\,dex) at a given age
as seen from the panel with \feh\ in Fig. \ref{fig:abun-age}.

The \xfe -age slopes given in Table \ref{table:fits} agree remarkably well
with the slopes determined by \citet{spina16a} for a sample of
13 solar twins with ages less than 6\,Gyr. For 14 elements, the $b$-coefficients
agree within 1-sigma and for two elements (O and Ca) within 2-sigma.
Only in the case of Mn, there appears to be a significant difference.
Spina et al. obtain a steeper slope, $b = 18.4\,\pm 3.2 \, 10^{-3}$\,dex Gyr$^{-1}$,
in comparison with $b = 7.2\,\pm 2.0 \, 10^{-3}$\,dex Gyr$^{-1}$ in this work. 

It is also interesting to compare with the age trends
obtained by \citet{dasilva12} for a sample of 25 solar-type stars
having 5500\,K $< \teff < 6100$\,K and $-0.3 < \feh < +0.3$ for which
they derived \xfe\ abundance ratios with errors ranging from 0.03 to 0.07\,dex.
They divided the stars into two groups, younger and older than the Sun,
and looked for correlations between \xfe\ and age for each group.
Slope coefficients are not given, but from Fig. 11 in  
\citet{dasilva12} it is seen that the trends agree qualitatively 
with the trends determined in the present paper. \mgfe , \sife , and \znfe\ increase
with age for both age groups, whereas \cafe\ and \crfe\ are nearly constant. \cfe , \nafe , and
\nife\ increase with age for the young group. 
\yfe\ and \bafe\ decrease with increasing age, but \bafe\ flattens out
for the old stars like in Fig. \ref{fig:bay-age} of this paper. Furthermore,
\mnfe\ increases with age for the young group but decreases for 
for the old group as also found in present paper. Altogether, the 
results of \citet{dasilva12} suggest that the age trends found for solar twins
are shared by solar-type stars with a larger range of metallicities.

Furthermore, \citet{spina16b} have recently determined
new high-precision abundances and ages for nine solar twins and discussed the
trends of abundance ratios with age for a total of 41 solar twins having thin-disk 
kinematic and ages between 1 and 8\,Gyr (on the Yonsei-Yale age scale)
including the stars in \citet{spina16a} and Paper\,I.
The \xfe -age results agree well with those shown
in Fig. \ref{fig:abun-age} of the present paper, but 
the data are fitted by hyperbolic functions, which in the cases of
\nafe\ and \nife\ have a maximum around 5\,Gyr. According to this interpretation
\nafe\ and \nife\ increase from 1 to 4\,Gyr but decrease from 6 to 8\,Gyr.
This is an interesting interpretation of the data,
but it remains to be seen if a maximum in \nafe\ and \nife\
at $\sim$5\,Gyr can be explained by chemical evolution models.

\subsection{\xfe \,-\,\Tc\ trends}
\label{Tc.trends}
It is clear from  Fig. \ref{fig:abun-age} and the discussion in 
Sect. \ref{age-correlations} that variations in abundance ratios among solar twins
younger than 6\,Gyr are mainly due to differences in age. Still, it is interesting
to investigate if there is a correlation between the residuals in 
the \xfe -age relations, i.e., $\xfe_{\rm res} = \xfe  - (a + b \,\cdot\,Age$),
and elemental condensation temperature \citep{lodders03}. 
As mentioned in Sect. \ref{age-correlations}, the mean Galactocentric
orbital distance, $R_{\rm m}$, of these stars are about the same, so a 
possible dependence of the \Tc -slope on $R_{\rm m}$
\citep{adibekyan14, adibekyan16} is not an issue.
  
\begin{figure*}
\resizebox{\hsize}{!}{\includegraphics{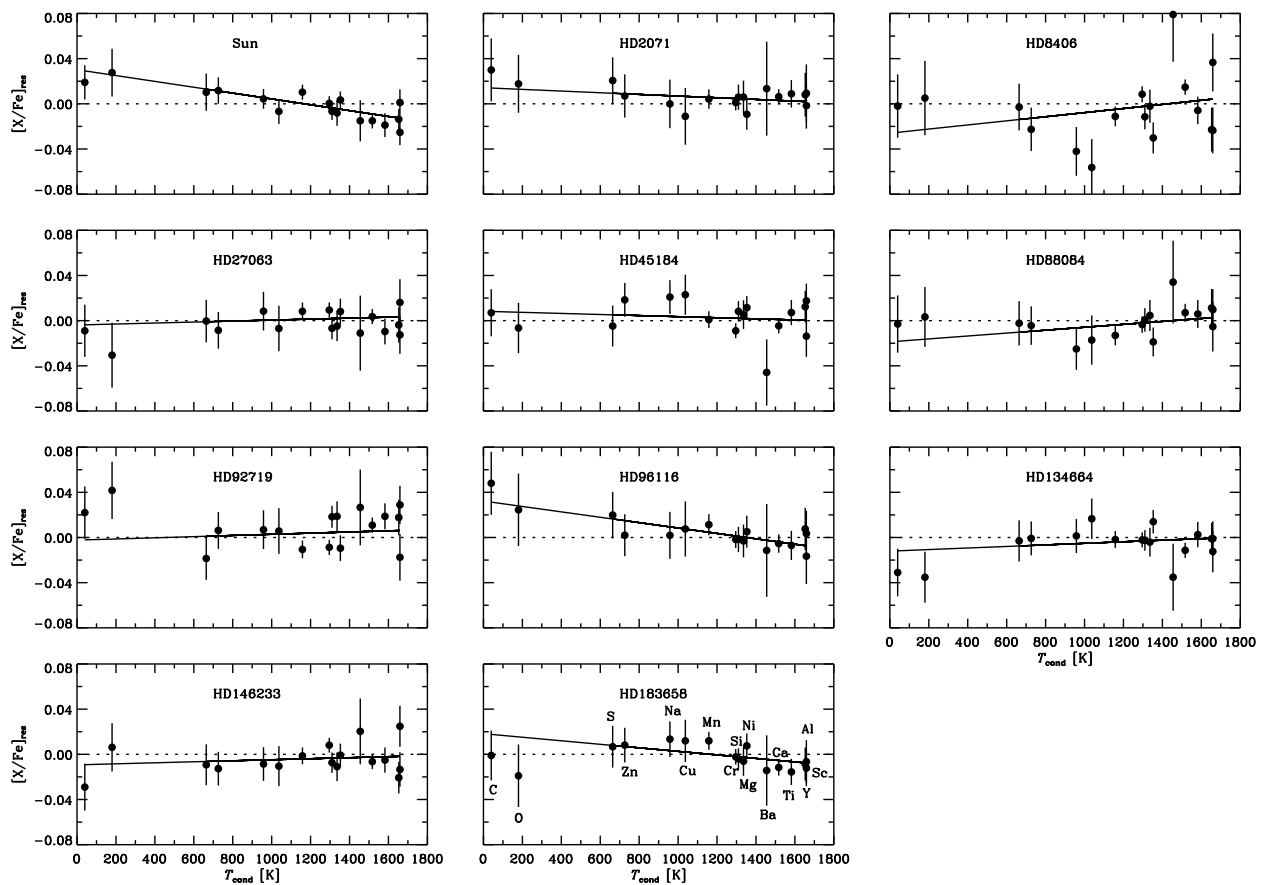}}
\caption{Abundance ratios corrected for Galactic evolution as a function of
elemental condensation temperature for the Sun and ten solar twins with ages less than 6 Gyr.
Element designations are shown in the panel for HD\,183658.}
\label{fig:element-Tc}
\end{figure*}

Fig. \ref{fig:element-Tc}  shows $\xfe_{\rm res}$ as a function of
\Tc\ for the Sun and  stars younger than 6\,Gyr. 
The lines show linear fits, $\xfe_{\rm res} = c + d \,\cdot \, \Tc$,
obtained  by weighting each point by the inverse square of the error of 
$\xfe_{\rm res}$ calculated by adding $\sigma \xfe$,
$b \cdot \sigma (Age)$, and the error of the \xfe \,-\,age fit at the age
of the star in quadrature. Y and Ba are included in the fits,
but due to the relatively large error of $\xfe_{\rm res}$ for these
two elements, they have only a small effect on the fit.

The coefficients of the $\xfe_{\rm res}$\,-\,\Tc\ fits and their 1-sigma errors
are given in Table \ref{table:Tc_slopes}. As seen,
only the Sun and \object{HD\,96116} show a significant trend of 
$\xfe_{\rm res}$ as a function of \Tc . For the other stars, the 
$\xfe_{\rm res}$\,-\,\Tc\ slope deviates less than 2-sigma from zero slope.
For the same stars, the (uncorrected) \xfe - \Tc\  slope ranges from zero 
to $7 \cdot 10^{-5}$dex\,K$^{-1}$ and depends on stellar age (see Fig. 14 in Paper\,I).
This shows the importance of correcting for Galactic evolution before
discussing correlations between \xfe\ and \Tc .

Interestingly, the Sun\,\footnote{The errors of $\xfe_{\rm res}$ for the Sun 
are somewhat smaller than the corresponding errors for the stars 
because the $b \cdot \sigma (Age)$ error contribution is negligible for the Sun.}
has the most significant variation of $\xfe_{\rm res}$
with \Tc , i.e.,  a slope of $-2.6 \, \pm 0.7 \cdot  10^{-5}$dex\,K$^{-1}$.
Thus, it seems that the Sun has an exceptional high ratio between volatile and refractory 
elements when compared to solar twins as first found by \citet{melendez09}. 
They did not correct for Galactic evolution of abundance ratios,
but the mean age of their sample of 11 solar twins was estimated to be
4.1\,Gyr, i.e. close to 
the age of the Sun. The variation of \xfe\ over a \Tc\ range of 1800\,K 
was found to be $\sim \! 0.08$\,dex compared to $\sim \! 0.05$\,dex in this paper.
An even smaller \Tc -variation, $\sim \! 0.04$\,dex, was found by 
\citet{spina16a}, when comparing the Sun to 13 solar twins (including corrections 
for chemical evolution). These different results may arise because the Sun is 
compared to different samples of solar twins.
For the sample in this paper, only one star (\object{HD\,96116})
out of 10 has a negative $\xfe_{\rm res}$\,-\,\Tc\ slope like the Sun, whereas the 
slopes for the other stars are within  $\pm 1.8 \cdot  10^{-5}$dex\,K$^{-1}$.
In contrast, \citet{spina16a} found evidence of a larger range of \Tc -slopes, 
$\pm 4 \cdot  10^{-5}$dex\,K$^{-1}$.  Obviously, a larger sample of solar twins is needed
to obtain better information on the distribution of the $\xfe_{\rm res}$\,-\,\Tc\ slopes 
to see how exceptional the Sun is.

\begin{table}
\caption[ ]{Linear fits, $\xfe_{\rm res} = c + d \cdot \Tc$.}
\label{table:Tc_slopes}
\centering
\setlength{\tabcolsep}{0.10cm}
\begin{tabular}{lccc}
\noalign{\smallskip}
\hline\hline
\noalign{\smallskip}
  Star & $c$ & $d$ & $\sigma \xfe_{\rm res}$  \\
       & [dex] &  [$10^{-5}$dex\,K$^{-1}$] & [dex] \\
\noalign{\smallskip}
\hline
\noalign{\smallskip}
         Sun & $+0.030\; \pm 0.009$ & $-2.6\; \pm 0.7$ & 0.008  \\
    HD\,2071 & $+0.014\; \pm 0.014$ & $-0.7\; \pm 1.1$ & 0.009  \\
    HD\,8406 & $-0.026\; \pm 0.015$ & $+1.8\; \pm 1.1$ & 0.032  \\
   HD\,27063 & $-0.004\; \pm 0.013$ & $+0.4\; \pm 1.0$ & 0.012  \\
   HD\,45184 & $+0.008\; \pm 0.011$ & $-0.5\; \pm 0.9$ & 0.017  \\
   HD\,88084 & $-0.018\; \pm 0.013$ & $+1.3\; \pm 1.0$ & 0.014  \\
   HD\,92717 & $-0.002\; \pm 0.012$ & $+0.5\; \pm 0.9$ & 0.020  \\
   HD\,96116 & $+0.032\; \pm 0.015$ & $-2.4\; \pm 1.1$ & 0.009  \\
  HD\,134614 & $-0.012\; \pm 0.012$ & $+0.7\; \pm 0.9$ & 0.014  \\
  HD\,146233 & $-0.009\; \pm 0.011$ & $+0.5\; \pm 0.9$ & 0.013  \\
  HD\,183658 & $+0.018\; \pm 0.012$ & $-1.6\; \pm 0.9$ & 0.012  \\
\noalign{\smallskip}
\hline
\end{tabular}

\end{table}

\subsection{Nucleosynthesis and chemical evolution}
\label{nucleosyn}
As discussed in Paper\,I, the decreasing trend of \xfe\ with decreasing stellar age
for elements made in massive stars
is probably due to an increasing contribution of Fe from Type Ia SNe
in the course of time. The flat distribution of \cafe\ is a problem,
but may be explained if low luminosity SNe synthesizing large amounts of Ca
\citep{perets10} are contributing to the chemical evolution of Ca
\citep{mulchaey14}. \crfe\ is also nearly constant in time, indicating 
that the ratio of contributions from Type II and Ia SNe 
is about the same for Cr and Fe. Furthermore,  
the steep increase of \yfe\ with
decreasing age is probably due to the delayed contribution of $s$-process elements
from low-mass AGB stars.
In the following, it is discussed how the elements analyzed in this paper
(Sc, Mn, Cu, and Ba) fit into these interpretations of the \xfe -age trends. 

\subsubsection{Scandium}
\label{scandium}
As reviewed by \citet{romano10}, Sc is made in core-collapse (Type II) SNe
during neon burning and explosive oxygen and silicon burning \citep{woosley95}.
Type Ia SNe give a negligible contribution to Sc according to \citet{iwamoto99}.
Calculated yields \citep[e.g.,][]{timmes95, kobayashi06} lead to too low
Sc/Fe ratios compared to observed values, but this problem may be solved
by including neutrino interactions \citep{frohlich06, yoshida08}. 

\begin{figure}
\resizebox{\hsize}{!}{\includegraphics{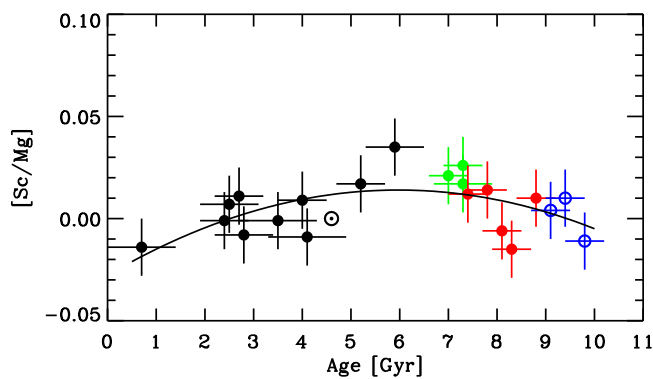}}
\caption{\scmg\ as a function of stellar age with the
same symbols as in Fig. \ref{fig:abun-age}.
The line shows a quadratic least squares fit to the data with all stars and
the Sun included.}
\label{fig:scmg-age}
\end{figure}

It has been much discussed if Sc follows Fe, i.e., $\scfe \sim 0$ at all metallicities
\citep{gratton91, prochaska00a} or if Sc behaves like an $\alpha$-capture element,
i.e., $\scfe \! \sim \!  0.2$ in metal-poor halo and thick-disk stars
\citep{zhao90, nissen00, reddy06, adibekyan12, battistini15}. 
The solar twin data support the second possibility as seen from Fig. \ref{fig:scmg-age}.
Although there is a rise of \scmg\ with increasing age for stars
younger than 6\,Gyr and a corresponding decline with age for the older stars,
the total variation of \scmg\ is not greater than 0.05\,dex and the 
various groups of stars agree  well with the fitted quadratic relation.
In contrast, \scfe\ shows a total variation close to 0.2\,dex 
that cannot be well fitted by a simple function over the whole age
range from 1 to 10\,Gyr.

\subsubsection{Manganese}
\label{manganese}
Mn can be made by explosive silicon burning and nuclear statistical processes
in Type II SNe \citep{woosley95} and in Type Ia SNe \citep{iwamoto99}.
Calculated yields are uncertain, because they depend on assumptions related to 
the explosion mechanism, mass cut and number of electrons per nucleon
\citep{thielemann96, mishenina15a}. According to the yield calculations of 
\citet{kobayashi06}, the Type Ia to Type II yield ratio is about three times
higher than the corresponding ratio for Fe. 
Since long, this has been considered as an explanation of
the increasing trend of \mnfe\ with increasing \feh\ found
from LTE analysis of \MnI\ lines in spectra of halo and disk stars
\citep{reddy06, feltzing07, adibekyan12, battistini15, mishenina15a}. 
In addition, \citet{cescutti08} have argued that in order to explain the
different \mnfe \,-\,\feh\ trends in the Galactic bulge, the solar neighborhood,
and the Sagittarius dwarf galaxy, one has to invoke metallicity dependent
Type Ia SNe yields.

Interestingly, \mnfe\ in the solar twins vary very little with age
as seen from Fig. \ref{fig:abun-age}. For stars older than 6\,Gyr,
there is a slight increase in \mnfe\ with decreasing age as one would
expect if Type Ia SNe are more important contributors to Mn
than to Fe. However, at ages less than 6\,Gyr, \mnfe\
decreases slightly as time goes on. This does not agree with a
rize of \mnfe\ of about 0.3\,dex, when \feh\ increases from $-1$ to
solar metallicity, as found when analyzing  \MnI\ lines in LTE.
The solar twin results are in better agreement with the near-constancy
of \mnfe\ as a function of \feh\ derived by \citet{bergemann08} and
\citet{battistini15}, when non-LTE effects are taken into account. 
At $\feh \sim -1$, the non-LTE correction for F and G dwarfs amounts to
approximately +0.3\,dex.

\subsubsection{Copper}
\label{copper}
Two processes have been suggested for the nucleosynthesis of Cu
in massive stars: explosive Ne-burning \citep{woosley95} 
and the weak $s$-process in which neutrons from the $^{22}$Ne\,$(\alpha , n)\,^{25}$Mg 
reaction are added to iron-seed nuclei \citep[e.g.,][]{raiteri93}.
Furthermore, \citet{matteucci93} argued that a substantial fraction of Cu
has to be made in Type Ia SNe in order to explain the increase of \cufe\ with
metallicity, but this is not supported by yield calculations
\citep{iwamoto99, kobayashi06}. More recently, \citet{romano07} showed that
\cufe\ versus \feh\ for stars in the solar neighborhood and in
$\omega$\,Cen \citep{cunha02, pancino02} can be explained by metallicity
dependent yields of massive stars without invoking any contribution
from Type Ia SNe.

\begin{figure}
\resizebox{\hsize}{!}{\includegraphics{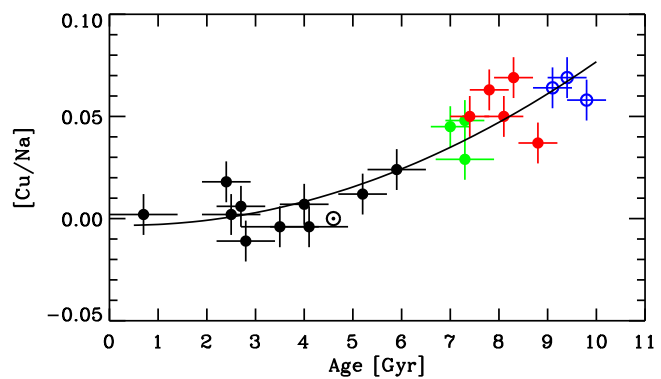}}
\caption{\cuna\ as a function of stellar age with the
same symbols as in Fig. \ref{fig:abun-age}.
The line shows a quadratic least squares fit to the data with all stars and
the Sun included.}
\label{fig:cuna-age}
\end{figure}

As seen from  Fig. \ref{fig:abun-age}, the pattern  of \cufe\ versus age looks
very much the same as the pattern of \nafe , and the slope coefficients for
stars younger than 6\,Gyr agree within the errors, i.e., $b = 0.025 \pm 0.003$\,dex Gyr$^{-1}$
for \nafe\ and $b = 0.029 \pm 0.004$\,dex Gyr$^{-1}$  for \cufe .
This is also seen 
in Fig. \ref{fig:cuna-age}, where \cuna\ is plotted as a function of stellar age.
\cuna\ is nearly constant for ages up to 6\,Gyr and although the ratio
rises somewhat at older ages, high- and low-\nafe\ stars 
(marked in green and red, respectively) as well as the thick-disk stars (marked in blue)
follow the same \cuna -age relation. This points to a close connection between
the nucleosynthesis of Na and Cu. Since the differences of \nafe\
among the older stars are probably due to differences in the neutron
excess arising from the $^{22}$Ne\,$(\alpha , n)\,^{25}$Mg reaction
during hydrostatic carbon burning in massive stars, 
the tight \cuna -relation suggests that the yield of Cu also depends on the 
neutron excess, i.e., Cu is primarily made by the weak $s$-process.

\subsubsection{Barium}
\label{barium}
In Galactic disk stars, Ba (like Y, Zr, and La) is 
primarily made  by the main $s$-process during
shell He-burning in AGB stars with a minor (10 - 20\,\%) contribution
from the $r$-process \citep{arlandini99}.
\bafe\ does not vary significantly as a function of \feh\
\citep[e.g.,][]{reddy06, bensby14},
but \citet{edvardsson93}, \citet{bensby07}, and \citet{dasilva12}
found evidence that \bafe\ increases with
decreasing age for solar metallicity stars. For open
clusters and associations, there is also evidence for an increase of
\bafe\ with decreasing age \citep{dorazi09, jacobson13, mishenina15b},
whereas a corresponding increase of \yfe , \zrfe , and \lafe\ was not
found. This is puzzling, because these $s$-process elements are 
expected to behave in the same way as Ba, and  the question has been raised
if the Ba abundances derived from the rather strong \BaII\ lines
in spectra of young stars can be trusted \citep{reddy15}.
On the other hand, \citet{maiorca11} did find an increase of \yfe , \zrfe .
and \lafe\ towards younger age of clusters, and suggested that this 
(and the \bafe\ increase) can be explained if the $s$-process yields
of AGB stars with masses between 1 and 1.5\,$M_{\odot}$ are enhanced by
a factor of $\sim \! 4$ \citep{maiorca12} relative to the yields of \citet{busso01}.
  
As seen from Fig. \ref{fig:bay-age}, both \yfe\ and \bafe\ in solar twins
show a very clear increasing trend with decreasing stellar age.
This indicates that the over-abundance of Ba/Fe derived for young
open clusters can be trusted and supports the trend of Y abundances found 
by \citet{maiorca11}. There are, however, interesting differences in
the \yfe\ and \bafe\ trends for the solar twins. While \yfe\ continues
to decrease with age for the oldest stars, \bafe\ flattens out at a 
level of about $-0.03$\,dex for stars older than 6\,Gyr. Furthermore,
the increase of \bafe\ with decreasing age for stars younger than 6\,Gyr
is steeper than the corresponding increase of \yfe . Assuming that
the increase is due to the delayed contribution of $s$-process
elements from low-mass AGB stars, the different age trends of \yfe\ and \bafe\
suggest that the Ba/Y yield ratio increases with decreasing stellar mass. 
This agrees with recent calculations of AGB yields
for solar metallicity models by \citet{karakas16}. 
According to their Fig. 15, the ratio between heavy $s$-process
elements (Ba, La, and Ce) and light $s$-process elements (Sr, Y, and Zr)
increases by $\sim 0.3$\,dex when the initial mass of the
AGB models decreases from 3 to 1.5\,$M_{\odot}$. 

\subsection{Chemical clocks}
\label{clocks}
\begin{figure}
\resizebox{\hsize}{!}{\includegraphics{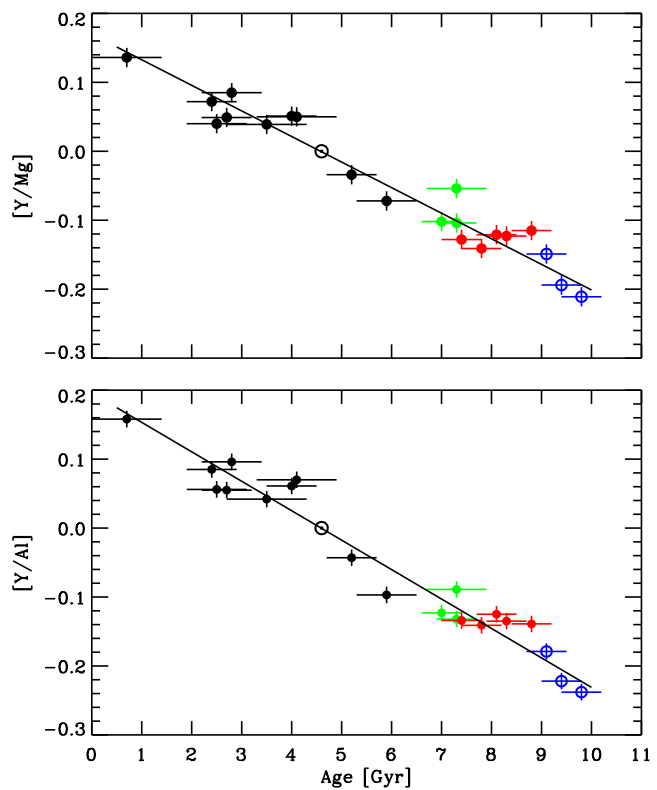}}
\caption{\ymg\ and \yal\ versus stellar age with the
same symbols as in Fig. \ref{fig:abun-age}.
The lines show the linear fits given in Eqs. (5) and (6).}
\label{fig:ymgal-age}
\end{figure}

As mentioned in Paper\,I, there is a tight linear relation between
\ymg\ and age of the solar twins. Fig. \ref{fig:ymgal-age} shows that
the  high- and low-\nafe\ stars
as well as the \alphafe -enhanced stars fit the relation very well.
A maximum likelihood fit (including all stars and the Sun) with
errors in both coordinates taken into account gives
\begin{eqnarray}
\ymg = 0.170 \, (\pm 0.009) - 0.0371 \, (\pm 0.0013) \,\,Age \,{\rm [Gyr]} 
\end{eqnarray}
with $\chi^2_{\rm red} = 1.0$ and a rms scatter of 0.024\,dex in \ymg\
corresponding to a scatter of 0.6\,Gyr in age. 
This relation is valid for the ASTEC ages; using Eq. (1) to convert to the
Yonsei-Yale age scale, the slope coefficient becomes $-0.0419$\,dex Gyr$^{-1}$, i.e.,
close to the slope of $-0.0404$\,dex Gyr$^{-1}$ determined in Paper\,I,
where the three thick-disk stars and the Sun were not included in the fit.

Fig. \ref{fig:ymgal-age} shows that there is also a very good
correlation between \yal\ and stellar age:
\begin{eqnarray}
\yal = 0.196 \, (\pm 0.009) - 0.0427 \, (\pm 0.0014) \,\, Age \, {\rm [Gyr]}. 
\end{eqnarray}
Again, $\chi^2_{\rm red}$ is close to one and the scatter in \yal\ is 0.025\,dex.

\begin{figure}
\resizebox{\hsize}{!}{\includegraphics{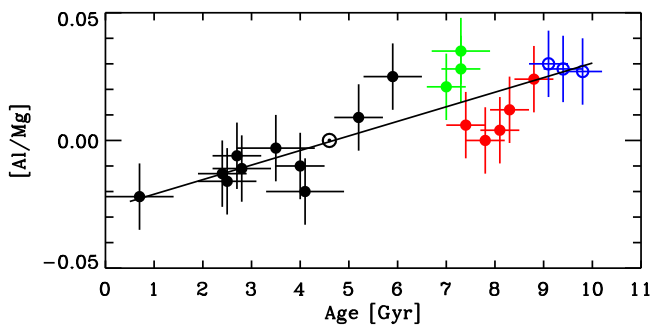}}
\caption{\almg\ versus stellar age with the
same symbols as in Fig. \ref{fig:abun-age}.
The line show the linear fit given in Eq. (7).}
\label{fig:almg-age}
\end{figure}

The fact that \ymg\ and \yal\ both have a tight linear dependence
on age means that \almg\ is closely correlated with age, as also 
seen from Fig. \ref{fig:almg-age}.
A linear fit to the data provides the relation
\begin{eqnarray}
\almg = -0.027 \, (\pm 0.005) + 0.0057 \, (\pm 0.0008) \,\, Age \, {\rm [Gyr]}. 
\end{eqnarray}
with a scatter of only 0.011\,dex in \almg . In particular,
the three $\alpha$-enhanced stars fit the relation very well
showing that Al behaves like an $\alpha$-element with a
slightly higher amplitude than Mg. 

The first evidence of a linear correlation between \ymg\ and stellar age 
with a slope similar to the value found in the present paper was 
presented by \citet{dasilva12} for a sample of 25 solar-type stars
albeit with a dispersion in \ymg\ on the order of $\pm 0.1$\,dex,
which is probably due to errors in the abundance determinations.
Recently, \citet{tuccimaia16} has determined high-precision
Y and Mg abundances for a sample of 88 solar twins and analogs for which 
\citet{ramirez14} determined stellar parameters and ages.  
The close correlation between \ymg\ and stellar age is confirmed for
this larger sample of stars in the solar neighborhood. 
The slope is about the same as in Eq. (2) but the dispersion in
\ymg\ ($\pm 0.037$\,dex) is larger than found in the present paper ($\pm 0.024$\,dex).

\section{Summary and conclusions}
\label{summary}
Abundances of Sc, Mn, Cu, and Ba have been determined for a sample of 21
solar twin stars with precisions around 0.01\,dex relative to the solar abundances.
The data are discussed together with high-precision abundances
of C, O, Na, Mg, Al, Si, S, Ca, Ti, Cr, Fe, Ni, Zn, and Y derived
in Paper\,I and used to obtain new information about the nucleosynthetic history
of elements in the Galactic disk from trends of abundance ratios with stellar age.

Stellar ages with precisions ranging from 0.4 to 0.8\,Gyr
were derived by comparing the position of stars in the \teff \,-\,\logg\ diagram
with isochrones based on stellar models calculated with the 
Aarhus Stellar Evolution Code \citep{jcd08}.
These  ages are linearly correlated with the ages based on Yonsei-Yale  isochrones
\citep{yi01, kim02}  derived in Paper\,I, but the ASTEC ages are about 10\,\% higher 
than the YY ages for the oldest stars. This difference of the age scale may
be due to the inclusion of metal diffusion in the ASTEC models, which effect is  
neglected in the YY models.  

For stars younger than 6\,Gyr and for all elements, \xfe\ as a function
of age can be fitted by linear relations having $\chi^2_{\rm red} \simeq 1$, and
for most stars, the residuals do not depend significantly on elemental condensation
temperature, \Tc . The Sun and HD\,96116 are exceptions, but in these cases 
the total amplitude in
the \Tc\ dependence is only about 0.05\,dex. For stars older than 6\,Gyr,
there is no clear dependence between \xfe\ and age. Three stars
with ages between 9 and 10\,Gyr have enhanced \alphafe\ ratios and
probably belong to the thick-disk population. 
Stars with ages between 6 and 9\,Gyr tend to split
up into two groups with high and low values of \xfe\ for
the odd-$Z$ elements Na, Al, Sc, and Cu.
The Type II SNe yield for these elements depends on the
neutron excess, and stars older than 6\,Gyr may therefore have been formed
in regions following different chemical evolution paths, whereas
stars younger than 6\,Gyr (including the Sun) were formed from 
well-mixed gas. These conclusions are based 
on a small sample ($N = 21$) of solar twins, and should be further explored
by determining high-precision abundances and ages for larger samples.
In particular, it would be interesting to investigate if there are correlations
between abundance ratios and kinematics of the stars.

According to the discussion in Sect. \ref{nucleosyn},
the trends of abundance ratios for Sc, Mn, Cu, and Ba with stellar age
provide new information on nucleosynthesis in the Galactic disk. In summary: 
i) The smooth and relatively small variation 
of \scmg\ with age suggests that Sc is made in Type II SNe along with the
$\alpha$-capture elements; 
ii) the near-constancy of \mnfe\ with age indicates
that the Type Ia to Type II SNe yield ratio for Mn is similar to
the corresponding ratio for Fe;
iii) the small scatter in \cuna\ at all ages suggests that Cu is 
primarily made by the weak $s$-process in massive stars; iv) the steeper
increase of \bafe\ with decreasing stellar age compared to the increase of
\yfe\ suggests that the Ba/Y yield ratio of AGB stars
increases with decreasing stellar mass as recently predicted by
\citet{karakas16}.

Finally, \ymg\ and \yal\ were found to be sensitive indicators of 
stellar age, which can be explained by prompt production of Mg and Al
in Type II SNe and an increasing contribution of yttrium from low-mass AGB stars
as time goes on. For the present sample of 21 solar twins, \ymg\ and \yal\
depend linearly on age with a slope close to $-0.04$\,dex Gyr$^{-1}$. 
This shows that \ymg\ or \yal\ can be used as chemical clocks
to determine relative ages of solar metallicity stars with precisions 
of about one Gyr. The recent work by \citet{adibekyan16} suggests that the
\ymg \,-\,age relation is also valid outside the solar neighborhood and for 
metallicities somewhat different from solar.

\begin{acknowledgements}
Vardan Adibekyan is thanked for a constructive referee report that helped
to improve the paper significantly and Bengt Gustafsson for 
inspiring comments on a first version of the manuscript. 
Maria Bergemann is acknowledged for calculating non-LTE corrections for the 
the \MnI\ lines applied to derive Mn abundances and
J{\o}rgen Christensen-Dalsgaard for providing ASTEC 
evolutionary tracks for the mass-range of the solar twins. 
Funding for the Stellar Astrophysics Centre is provided by The
Danish National Research Foundation (Grant agreement no.: DNRF106).
The research is supported by the ASTERISK project
(ASTERoseismic Investigations with SONG and Kepler)
funded by the European Research Council (Grant agreement no.: 267864).
This research made use of the SIMBAD database operated
at CDS, Strasbourg, France.
\end{acknowledgements}

\bibliographystyle{aa9}
\bibliography{nissen}


\end{document}